# SANCTIONS AND IMPORTS OF ESSENTIAL GOODS:
# A CLOSER LOOK AT THE EQUIPO ANOVA (2021) RESULTS

Francisco Rodríguez[1]

Josef Korbel School of International Studies

University of Denver

**Abstract:** We revisit the results of Equipo Anova (2021), who claim to find evidence of an improvement in Venezuelan imports of food and medicines associated with the adoption of U.S. financial sanctions towards Venezuela in 2017. We show that their results are consequence of data coding errors and questionable methodological choices, including the use an unreasonable functional form that implies a counterfactual of negative imports in the absence of sanctions, the omission of data accounting for four-fifths of the country's food imports at the time of sanctions and incorrect application of regression discontinuity methods. Once these errors are corrected, the evidence of a significant improvement in the level and rate of change in imports of essentials disappears.

[1] Josef Korbel School of International Studies, University of Denver. E-mail: Francisco.Rodriguez4@du.edu. This paper was partly written while I was a visiting senior economist at the Center for Economic and Policy Research; I thank that institution for its support. I thank María Eugenia Boza, Nicolás Idrobo, Dorothy Kronick, Alex Main, Geoff Ramsey and Mark Weisbrot for comments and suggestions and Jacques Bentata, Giancarlo Bravo, Luisa García and Camille Rodríguez for excellent research assistance. The replication data for this paper can be found here. Although I am the single author of this paper and assume responsibility for any mistakes, I also wish to recognize that this research is the result of a collective undertaking enriched by the contributions, ideas and insights of colleagues, students and staff. To acknowledge and do justice to their contributions, I have written the paper using the first-person plural voice.

1. **Introduction**

How do economic sanctions affect living conditions in target countries? Concerns about the effects of sanctions on vulnerable populations have occupied a prominent place in the literature on economic statecraft. A growing literature has argued that sanctions have adverse effects on outcomes across a broad range of human development indicators ranging from per capita income (Neuenkirch and Neumeier, 2015; Spinter and Klomp, 2021), poverty (Neunenkirch and Neumeier, 2016), and inequality (Afersogbor and Mahadevan, 2016) to public health (Peksen, 2011; Petrescu, 2016; Allen and Letskian, 2012) and human rights (Wood, 2008; Peksen, 2011).

However, a recent study on Venezuela argues that restrictions on external financing imposed by the United States in 2017 are associated with an improvement in imports of essential goods (Equipo Anova, 2021). According to this study, the introduction of financial sanctions coincides with a discontinuous break in trend of key categories of imports of basic goods, leading to statistically significant increases in both the level and the rate of change of both food and medicines imports.

The findings of this study are remarkable in that they stand in stark contrast with those of the existing literature. Table 1 summarizes the results of 32 published studies that provide quantitative estimates of the effects of sanctions on living conditions in target countries. Of these, 30 studies find negative effects of sanctions on indicators of living standards in target countries and one finds ambiguous effects.[2] Equipo Anova (2021) provides a singular exception, reporting statistically significant improvements associated with the adoption of sanctions in a target country across all relevant indicators of living standards studied.[3] Furthermore, many of the country case studies written in the literature have emphasized restrictions on import capacity as being the key causal mechanism through which sanctions affect living conditions in target countries (Office of the UN Coordinator for Afghanistan, 2000; Kheirandish, Varahrami, Kebriaeezade and Cheraghali 2018; Oliveros, 2020; Rodríguez, 2021; Batmanghelidg, 2022; Carter Center 2022;), a channel of causation whose operation is directly contradicted by the results of the Equipo Anova (2021) study.

The purpose of this paper is to more closely examine the basis for the Equipo Anova (2021) result and to identify whether it effectively provides a counterexample to the existing literature. We focus on the sensitivity of the results to choices in functional form, definition of the dependent variable, and statistical methods used to test for the existence of discontinuities.

---

[2] This table is a summary of Tables 1 and 2 of Rodríguez (2022b) which provides greater details on the methods and results of each of the studies. As discussed in that paper, some studies find a mixture of significant and insignificant results depending on the indicator and specification. Nevertheless, 30 of the published studies highlight the negative significant results as the main findings of their results, and one stresses the ambiguity of effects. The single exception is the Equipo Anova (2021) paper, in that it highlights as normatively significant only the indicators of imports of essential goods in which it finds positive effects.

[3] While Equipo Anova (2021) does find negative effects of sanctions on oil production, their discussion suggests that this effect can be disregarded due to their finding of improvement in imports of essentials: "even though the strategy of sanctions against [state-owned oil company] PDVSA could be responsable - only partial [sic]-of the decline observed in oil production and, with that, of the fiscal and external revenues of the Venezuelan economy, there is no evidence that sanctions have had a negative effect in the availability of basic humanitarian inputs."(2021, p. 2)

# Table 1: Summary of cross-country panel data and country-level studies.

| Authors | Outcome variable | Results | Data coverage | Authors | Outcome variable | Results | Data coverage |
|---|---|---|---|---|---|---|---|
| Hufbauer et al, (1985, 1990, 2007) | Economic cost to target countries | Average cost to target of comprehensive sanctions regimes is 4.2% of GDP | Cross-country | Dai et al (2021) | Trade flows between countries | The imposition of sanctions leads to a 77-82% decline in bilateral trade. | Cross-country |
| Wood (2008) | Human Rights Abuses | The most severe UN sanctions lead to an increase in the probability of repression from 5% to 25%; for U.S. sanctions the increase is to 16%. | Cross-country | Ha and Nam (2021) | Life expectancy | The imposition of economic sanctions reduces avarage life expectancy by 0.3 years. The effect is present only for trade and other sanctions. Countries with more developed financial systems and institutions are better able to alleviate the effect of sanctions on health. | Cross-country |
| Peksen (2009) | Physical integrity rights of citizens | Economic sanctions lead to significant increases in human rights violations in all measures of physical intergity considered as well as in aggregate indices | Cross-country | Splinter and Klomp (2021) | Growth collapses | The likelihood of experiencing a growth collapse rises by 9% in the first three years after sanctions are imposed | Cross-country |
| Peksen and Cooper (2010) | Democracy | Sanctions lead to a 7% reduction in the average Freedom House democracy score the year after sanctions are imposed, an effect that rises to 16% in the case of extensive sanctions | Cross-country | Gutmann et al. (2021) | The Economic Effects of International Sanctions: An Event Study | International sanctions slow GDP growth in the first (2.2 pp) and second (1.8 pp) years of a sanctions episode. The effect seems to be mostly caused by US unilateral sanctions and financial sanctions, and operates through a decline in Foreign direct investment, which drops by 39% in the first year of an episode of sanctions. | Cross-country |
| Peksen (2011) | Under-five child mortality rates | A one-standard deviation increase in the cost of sanctions leads to 4 percent increase in mortality. U.S. sanctions leads to a 35 percent increase in mortality. | Cross-country | Van Bergeijk (2015) | Political outcomes, GDP, oil and gas rents, government consumption, imports | Sanctions lead to a decline in GDP that is statistically significant only in the short term. Sanctions have no statistically significant effect on democracy either in the short nor long term. | Iran |
| Allen and Lektzian (2012) | Public health | Sanctions that have a large economic effect on target states can have severe public health consequences. These consequences are substantively similar to those associated with major military conflicts. | Cross-country | Farzanegan et al. (2016) | Macroeconomic Variables and Household welfare | Sanctions lead to a decline in total imports by 20%, total exports by 16.5%, private consumption by 3.9%, capital income by 3.8% and GDP by 2.2% | Iran |
| Choi and Lio (2013) | Incindents of international terrorism | The imposition of sanctions leads to a 93% increase in incidents of international terrorism | Cross-country | Warburton (2016) | GDP Growth, Exports/GDP, Inflation. | Sanctions detrimentally affect the target's macroeconomic performance | Iran, Liberia, Rwanda, Sierra Leone |
| Neuenkirch and Neumeier (2015) | Economic growth | UN sanctions cause 2.0% decline in growth at time of sanctions, rising to a cumulative 26% over 10 years. U.S. sanctions lower growth by 0.9% and cumulative 13%. | Cross-country | Gharehgozlia (2017) | Real GDP | Sanctions caused a 17.3 percent decline in real GDP in 3 years. The higest effect of the sanctions took place 2012, with a 12.0 percent drop. | Iran |
| Afesorgbor and Mahadevan (2016) | Income Inequality | A sanctions episode increases the Gini coefficient by 1.7 points, while each additional year of sanctions adds 0.3 points to the Gini | Cross-country | Parker et al. (2017) | Child mortality | The authors focused on sanctions designed to stop human rights violations through 3T mines in DRC. They found that infant mortality rates rose in villages that depend economically on armed groups. Also, they found that infant mortality rates increased in villages that depend economically on "conflict-free" minerals. | Five eastern Congo provinces |
| Lucena and Apolinário (2016) | Human Rights Abuses | Protection against loss of life and torture is 1.7 times more likely to worsen under targeted sanctions han under no sanctions. The effect of targeted sanctions is not different from that of conventional sanctions. | Cross-country | Kholodilin and Netšunajev (2018) | GDP Growth | Sanctions had a weak negative effect on the growth rate of Russian GDP. | Russia and euro area. |
| Petrescu (2016) | Weight, height and probability of death of children | Being exposed to sanctions during the whole duration of a pregnancy leads to a decrease of .07 standard deviations in a child's weight | Cross-country | Rodríguez (2019) | Venezuelan oil production | Financial sanctions are associated with losses in oil revenues of USD 16.9 billion per year | Venezuela |
| Neuenkirch and Neumeier (2016) | Poverty gap | U.S. sanctions lead to increases in the poverty gap by 3.8 % of GDP. The effect grows to 7.9% for most severe sanctions and is reinforced with multilateral support. | Cross-country | Felbermayr et al. (2020) | International trade | Sanctions reduce trade with the target, with an effect that is economically significant and heterogeneous across countries and sectors. | Iran |
| Gutmann et al. (2017) | Life expectancy for males and females | UN sanctions are associated with a decrease in life expectancy of 1.2 years for men and 1.4 years for women, while U.S. sanctions are associated with a smaller decline of 0.4 years for men and 0.5 years for women | Cross-country | Equipo ANOVA (2021) | Essential goods imports and oil production | Financial sanctions are associated with an improvement in imports of essential goods but an acceleration in the rate of decline in oil production. | Venezuela |
| Gutmann et al (2018) | Economic sanctions and human rights: Quantifying the legal proportionality principle | US economic sanctions are associated with a deterioration of political rights but an improvement in women's emancipatory rights. The effect on emancipatory rights comes from sanctions that are not targeted at human rights goals and that are imposed unilaterally by the US, while multilateral and human-rights targeted sanctions are not associated with improvements in women's rights but are associated with deteriorations in political rights. | Cross-country | Morteza (2021) | Real GDP | Sanctions caused a 12.5% fall in Iran's real GDP in the first year and a 19.1% decline 4 years after the application of the sanctions, while real GDP remained 5% lower than its counterfactual 2 years after the removal of sanctions. | Iran |
| Kim (2019) | HIV infections, aid-related deaths | Sanctions episodes lead to increase in the HIV infection rate of children by 2.5% and an increase in AIDS-related deaths by 1%. | Cross-country | Hejazi and Emamgholipour (2022) | Food prices, food security and dietary quality | Sanctions caused a significant increase in food prices. The share of urban and rural households that were prone to food insecurity increased from 8.8% and 25.2% to 11.2% and 29.2%, respectively, from 2017 to 2019. | Iran |
| Wen et al (2020) | Energy imports | Unilateral sanctions lead to increases in energy imports, as do U.S. sanctions, economic sanctions, and greater sanctions intensity. Plurilateral, EU and UN sanctions have no significant effects. | Cross-country | Rodríguez (2022) | Venezuelan oil output | Sanctions caused large losses in oil production among firms that had acces to finance prior to sanctions compared to a control group that lacked that access. The effect of sanctions explains around half of the output drop observed in those firms. | Venezuela |

## 2. Background

Over the past decade, several countries have imposed a broad array of sanctions on persons and entities associated with the government of Nicolás Maduro in Venezuela. While most countries, including the members of the European Union and Canada, have limited their actions to sanctions that target the economic transactions of specific actors associated with the Venezuelan government[4], the United States has imposed additional restrictions barring most types of financial and trade transactions that involve the Venezuelan government.

The U.S. first adopted sanctions against the Venezuelan government in August 2017, when the Trump administration barred the issuance of new debt and the transfer of dividends from offshore entities to the Venezuelan government. In January 2019, the U.S. designated the state-owned oil monopoly Petróleos de Venezuela (PDVSA) as a specially designated national entity (SDN), thus imposing an embargo on all oil trade with Venezuela. In the same month, it recognized then National Assembly president Juan Guaidó as interim president of Venezuela, transferring to his appointees the management of Venezuela's offshore assets and the legal capacity to represent the Venezuelan state before US courts. In February of 2020, the U.S. began imposing secondary sanctions on Russian and Mexican companies that had helped sell Venezuelan oil in non-U.S. markets.

Between 2012 and 2020, Venezuela underwent the largest economic contraction documented in Latin American economic history since 1950, with per capita income falling by 72% (Rodríguez, 2022a). The extent to which this decline is affected by sanctions is a matter of intense debate. Most studies recognize that poor economic policies played a significant role in the onset of the contraction, as they left the economy unprepared to deal with the negative economic shock of the 2014 decline in oil prices. There is some discussion as to whether the 2017 financial sanctions can be associated with the ensuing decline in oil production; some authors (Rodríguez, 2018; Oliveros, 2021) suggest that the beak in trend oil production at the time is indicative of a decline while others point to alternative potential explanations (Bahar et al., 2019; Hausmann and Muci, 2019). Rodríguez (2022c) shows that oil joint ventures with access to finance prior to the 2017 financial sanctions suffered stronger declines in output than those that had already lost access to finance.

The worsening of socioeconomic indicators appears to be strongly related to declines in import capacity, with indicators of living standards improving consistently during the period of rising oil revenues (2000-2012) and deteriorating rapidly with the decline in export revenues and imports (2012-2020) (Rodríguez, 2022c). This is consistent with prior research on the Venezuelan economy, which finds that its economic fluctuations are largely driven by changes in oil export revenues (Rodríguez and Sachs, 1999; Hausmann and Rigobón, 2002, Rodríguez and Hausmann, 2012) as well as an extensive literature that shows that there are strong links between per capita income and non-income components of well-being (Pritchett and Summers, 1996; Filmer and Pritchett, 1999; UNDP, 2010).

---

[4] In this paper, we refer to the administration of Nicolás Maduro as the government of Venezuela. Some countries, however, recognize the interim government led by Juan Guaidó, president of the National Assembly elected in 2015, as the country's legitimate government since January of 2019. We follow the convention of referring to the authorities with *de facto* control of the territory as the government. This should not be interpreted as taking a position on the legitimacy of the claim to power of any of the parties involved.

### 3. The Equipo Anova (2021) study

A study published in January of 2021 by Venezuelan consultancy firm Anova (Equipo Anova, 2021) contends that the 2017 financial sanctions could have led to an increase and stabilization of imports of food and medicines. The argument is based on the finding of a break in trend in food and medicines imports at the time of the August 2017 sanctions. In the words of the study's lead author, "there is a strong temporal association between the start of the first economic sanctions in 2021 and the recovery of imports of humanitarian goods, in particular food and medicines" (Zambrano, 2022). In its conclusions, the report states that "it is possible to argue that the change in policy orientation of the government, which finally led to the flexibilization of the web of controls, was also an immediate consequence of the hardening of financial sanctions on PDVSA." (Equipo Anova, p. 3)

The Equipo Anova study has been widely reported in the local press and is often invoked in the country's policy debate (Amaya, 2021, Méndez, 2021, Polítika UCAB, 2021, Iturbe, 2021). The idea that sanctions have had a myriad of positive effects on the economy, particularly by forcing the government to enact policy reforms, is also often echoed by political actors. Opposition leader Juan Guaidó, who claims to be the country's legitimate president, said in a recent interview that sanctions had allowed the economy to experience a "dead cat bounce" in economic activity; in a recent speech, he added that "thanks to the sanctions today the [U.S.] dollar is used in Venezuela", alluding to the Maduro government's flexibilization of restrictions on the use of foreign currency in domestic transactions (Guaidó, 2022a, b). Legislator Armando Armas, who served as president of the Foreign Relations Committee of the opposition-controlled National Assembly elected in 2015, wrote that "to those who think the country is improving…I'll remind you that this is due to the policy of international sanctions on the regime and its collaborators." (Armas, 2022).

Let us consider in more detail the estimates presented by Equipo Anova. Formally, they estimate a simple linear trend model on a time series of Venezuelan import data ranging from April 2015 to December 2019:

$$Imports = \alpha_0 + \alpha_1 D + \alpha_2 t + \alpha_3 tD \qquad (1)$$

where $t$ denotes a time trend and $D$ is an indicator variable taking the value 1 on and after the adoption of sanctions on August 2017 and 0 before that date. Normalizing the time at the adoption of sanctions, $t_c$, to zero, they estimate equation (1) on medicines and food imports separately. They then test the hypotheses that $\alpha_1 = 0$ and $\alpha_3 = 0$ to evaluate respectively the change in level and rate of change of imports at the time of adoption of sanctions.

It is worth highlighting that Equipo Anova labels this research design as a regression discontinuity in time (RDiT) approach, using the term used to refer to time-series variants of the regression discontinuity design (RDD) framework (Hausman and Rapson, 2018). However, equation (1) differs significantly from what is typically understood in the literature as an RDD estimate. RDD estimation relies on the use of nonparametric regressions methods to estimate treatment effects on a subset of the data in the vicinity of a cutoff (Imbens and Lemieux, 2008; Lee and Lemieux, 2010; Hahn, Todd and Van der Klaauw, 2011 Lee, 2016). Typically, RDD studies use a variety of methods to select a window of data around the intervention of interest – also called a bandwidth – and use data inside that window, typically weighted by its proximity to the intervention date, to estimate treatment effects. Its implementation generally requires large data sets that contain many observations both before and after the intervention being evaluated.[5]

In contrast, Equipo Anova preselect a sample of 28 months before and after sanctions to carry out parametric estimation of (1) via ordinary least squares, presenting no argument as to why this would be an appropriate bandwidth choice. This choice is problematic because a key assumption of RDD is that the bandwidth is small enough to preclude major changes in other determinants of the dependent variable.[6] Yet the period over which Equipo Anova estimate regression (1) was also one during which major events with significant economic impacts took place, including a collapse in the price of the economy's main export, the overhaul of government food policies, reforms in price and exchange control regulations, the imposition of additional oil sanctions at periods different from the cutoff, and the holding of parliamentary, regional, constitutional and presidential elections, to name just a few potential confounding events.[7]

Aside from the debatable choice of estimation method and its interpretation, there are two additional major problems with the Equipo Anova estimates. One is their choice to use as a dependent variable a measure of import levels in US dollars, instead of the standard logarithmic specification used in the literature on macroeconomic time series. The other one is the omission of several import categories accounting for around four-fifths of the economy's food imports at the time of sanctions. We discuss each of these in turn.

When estimating equation (1), Equipo Anova specify their dependent variable as the level of imports measured in current US dollars. This implies that the pre- and post-sanctions parameters $\alpha_2$ and $\alpha_3$ will measure the change in absolute dollar amounts per period of time. In other words, a constant rate of decline in the Equipo Anova setup will not mean a constant rate of percentage decline, but rather a continued decline of *a specific amount of dollars per year*. Even if the percentage rate of change were to remain constant, changes in absolute dollar amounts would be captured by their estimation method as a break in trend.

Clearly, this decision strongly biases the method towards finding a break in trend in any setting in which there is a sustained decline (or increase) in the dependent variable. To see why, it's useful to think in terms of actual numbers. In 2014, Venezuela spent $7.5 billion in food imports; by 2017 those imports had fallen to $1.9 million. Thus, on average food imports declined by $1.8 billion per year (i.e., (7.5-1.9)/3). Specification (1) measured in levels implies that in the absence of a break in

---

[5] As discussed by Cunningham (2021, p. 252): "We need a lot of data *around* the discontinuities, which itself implies that the data sets useful for RDD are likely very large. In fact, large sample sizes are characteristic features of the RDD. This is also because in the face of strong trends in the running variable, sample-size requirements get even larger. Researchers are typically using administrative data or settings such as birth records where there are many observations" (emphasis in original).
[6] In the words of Hausman and Rapson (2018, p.535), the one reference cited by Equipo Anova on RD methods: "The use in RDiT of observations remote (in time) from the threshold is a substantial conceptual departure from the identifying assumptions used in a cross-sectional RD, and we show it can lead to bias resulting from unobservable confounders and/or the time-series properties of the data-generating process" (Hausman and Rapson, 2018, p. 535). For recent examples of RdiT applications and how they deal with issues of bandwidth choice, see Godard, Koning and Lindeboom (2019), Kim (2019), Carrears, Visconti and Acácio (2021), and Lovett and Xue (2021).
[7] Equipo Anova recognize that the adoption of price, exchange and import controls coincides with the post-sanctions period and that this fact clouds causal interpretation of the parameter. Despite this caveat, they offer as interpretation of their findings the possibility that the change in orientation of government policies is a consequence of the tightening of sanctions against PDVSA. (2021, p. 3).

trend, we should have expected those imports to continue declining at $1.8 bn a year over the following 28 months. That is clearly impossible, since imports cannot fall below zero.

**Figure 1: Implicit counterfactual trend in Equipo Anova estimates**

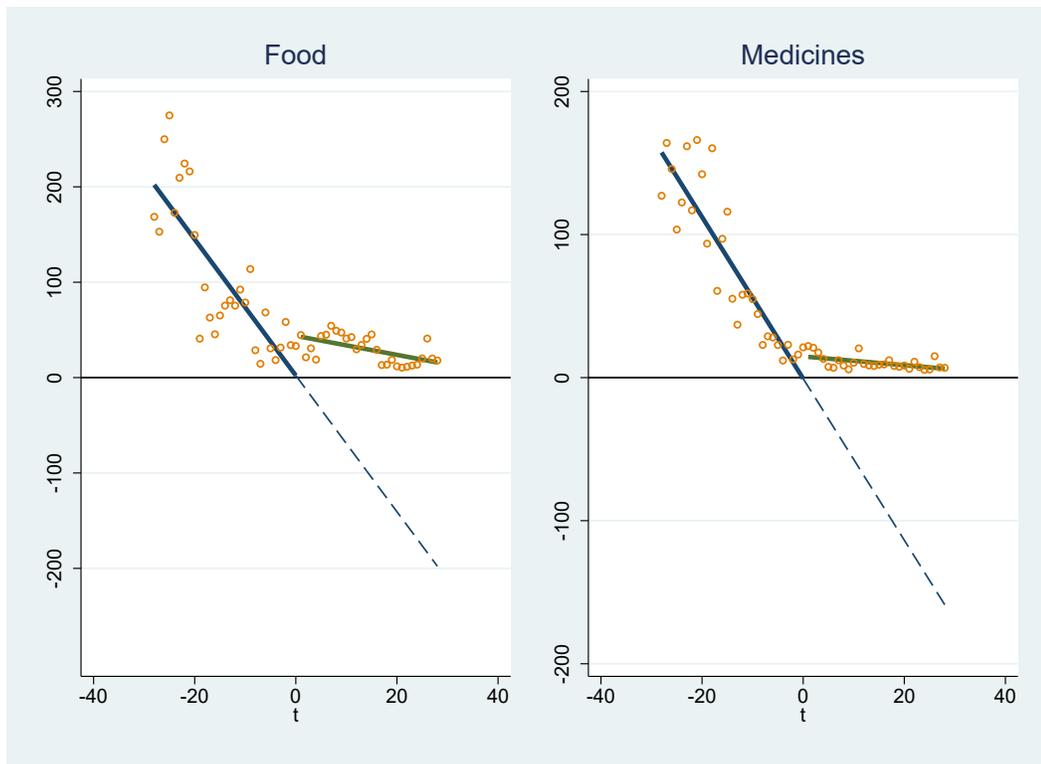

Sources: Equipo Anova (2021), UN Comtrade, own calculations

Figure 1 illustrates this point graphically. The figure reproduces the Equipo Anova estimates for both food and medicine imports yet expands the y-axis scale to allow us to display the counterfactual prediction of the evolution of imports in the absence of sanctions implied by their estimates.[8] The solid lines show the Equipo Anova estimates for both the pre- and post-sanctions periods, while the shaded line shows their projection of the pre-sanctions trend estimate to the post-sanctions period. Note that the Equipo Anova statistical tests for slope changes are essentially a comparison of the shaded line and the solid post-sanctions line. Yet this counterfactual is clearly inadequate as it implies that food imports would have fallen to around -$208mn per month and medicines imports to -$167mn per month by December of 2019 in the absence of sanctions.

The second major problem in the Equipo Anova results has to do with the omission of important categories of food items from the dependent variable for which they report results. Although Equipo Anova claim to use the same two-digit international Harmonised System categories from the UN Comtrade database as used in another recent study on Venezuelan food imports (Bahar et al., 2019), this is not true. The Equipo Anova study omits ten two-digit import categories from their indicator of food imports. These categories include cereals, mill products, oils and sugars, among others (for ease of exposition, we refer to these categories henceforth as "cereals and oils"). On the year of sanctions, these categories accounted for 79.7% of all food imports as calculated by Bahar et al.[9] In other words, the Equipo Anova results are based on a food imports series that covers just around one-fifth of food imports at the time of sanctions (Table 2).

---

[8] Equipo Anova did not respond to our request for their data set and code, so all our estimates are based on our reconstruction of their data, as explained in section 4 below.

**Table 2: Food categories used in Equipo Anova (2019) and Bahar et al (2019)**

| Code | Description | Bahar et al. (2019) | Equipo Anova (2021) | % of total imports (2017) |
|---|---|---|---|---|
| 02 | Meat | Included | Included | 3.2% |
| 03 | Fish | Included | Included | 0.2% |
| 04 | Dairy produce | Included | Included | 5.9% |
| 06 | Trees and other plants | Included | Included | 0.0% |
| 07 | Vegetables | Included | Included | 4.1% |
| 08 | Fruits and nuts | Included | Included | 0.5% |
| 10 | Cereals | Included | Excluded | 38.9% |
| 11 | Mill products | Included | Excluded | 5.5% |
| 12 | Oil seeds | Included | Excluded | 2.4% |
| 13 | Lac, gums, resins | Included | Excluded | 0.3% |
| 14 | Vegetable plaiting materials | Included | Excluded | 0.0% |
| 15 | Animal fats | Included | Excluded | 11.0% |
| 16 | Prepared meats | Included | Excluded | 3.5% |
| 17 | Sugars | Included | Excluded | 8.0% |
| 18 | Cocoa | Included | Excluded | 0.4% |
| 19 | Cereal preparations | Included | Excluded | 9.7% |
| 20 | Vegetable, fruit and plant preparations | Included | Included | 1.2% |
| 21 | Miscellaneous edible | Included | Included | 3.8% |
| 22 | Beverages, spirits and vinegar | Included | Included | 1.0% |
| 24 | Tobacco | Included | Included | 0.4% |

An additional issue with the Equipo Anova study refers to the availability of data for their period of study at the time of collection. Because Venezuela reports no disaggregated import data to the United Nations, the Comtrade estimates of the country's imports are based on disaggregated export data reported by Venezuela's trading partners. Since countries report and update this data to the UN with variable delays (generally up to 24 months), data collected too close to the time of study could be incomplete. Therefore, Equipo Anova was not using a complete data series of imports for the post-sanctions period at the time, but rather the series based on the subset of countries that had reported up to the end of the sample when their study was written at the end of 2020.[10]

---

[9] The omission of these categories appears to have been an inadvertent error caused by a typo in footnote 2 on page 6 of Bahar et al. (2019), which fails to mention that the authors included 1-digit Harmonised System code 1 in their calculations. Nevertheless, revision of the replication code published by Bahar et al. (2019) verifies that all code 1 categories were included in their study. Equipo Anova claim to have also constructed an alternate series including most of the omitted code 1 categories in their footnote 18 and to have obtained similar results to those reported in their note, yet do not publish these. Their claim is consistent with our findings because the problem in their results is caused by the combination of using incomplete data *and* a misspecified regression. Our reconstruction of their alternate measure finds that the trend break results hold only in the misspecified levels specification which they use in their paper, yet not in the appropriate logarithmic one.

As we show in Table 3, correcting for these omissions makes the Equipo Anova results on food imports disappear. [11] The columns of table 3 report results according to different specifications of the dependent variable, while the rows capture differences in functional form and econometric method. Thus, the first column reports the food import series used by Equipo Anova, which excludes cereals and oils, the second column reports the full Comtrade food import series that includes cereals and oils and the third column reports results using medicines imports. We highlight in bold face and shade the cells corresponding to the Equipo Anova published results in the first row, and those of the fully corrected specifications in the third row. Once we use the full Comtrade food imports series and use a logarithmic specification (column 2, panel 2), the change in slope for food imports is no longer significant while the change in levels is only significant at p=.094. Even that very weak result disappears once we use the updated data, driving both coefficients below standard significance levels (column 2, panel 3).[12] The contrast between the specification published by Equipo Anova and the logarithmic specification with complete and updated data is displayed graphically in Figure 2.

**Table 3: Trend Interruption Specifications**

| | | | Equipo Anova food imports (excluding cereals and oils) | Food imports (including cereals and oils) | Medicine imports |
|---|---|---|---|---|---|
| Levels | Data available as of October 2020 | Change in level | **44.86***<br>(13.82) | 91.94**<br>(42.62) | **19.6***<br>(5.70) |
| | | Change in slope | **6.46***<br>(1.09) | 6.21**<br>(2.79) | **5.47***<br>(0.49) |
| Logarithms | Data available as of October 2020 | Change in level | 0.53**<br>(0.25) | 0.38*<br>(0.22) | -0.05<br>(0.19) |
| | | Change in slope | 0.04***<br>(0.01) | 0.01<br>(0.01) | 0.07***<br>(0.01) |
| Logarithms | Data available as of December 2022 | Change in level | 0.56**<br>(0.26) | **0.34**<br>**(0.22)** | **-0.22**<br>**(0.20)** |
| | | Change in slope | 0.05***<br>(0.01) | **0.02**<br>**(0.01)** | **0.1***<br>**(0.01)** |

Specifications originally reported by Equipo Anova (2021)
Specifications that correct data omissions and functional misspecification using most recent data

---

[10] To be fair on this point, the Equipo Anova data was the best available data given their choice of window at writing. Yet given what we know about the delays in reporting, the data was clearly still preliminary at the moment and, as we show below, their results are additionally weakened when we use the updated data. A more prudent presentation of the results at the time should have stressed their preliminary nature.
[11] Note that by most recent data we refer to the updated data for the same period of study (April 2015-December 2019)
[12] In response to earlier critiques of the paper's functional form choice, lead Equipo Anova researcher Omar Zambrano posted on Twitter scatter plots showing results using a logarithmic specification and argued that they showed that the originally published results continued to hold even using a nonlinear functional form. However, the food results published on his Twitter thread are based on the restricted food import series that excludes cereals and oils and thus do not correct the data omission problem. See Zambrano (2021a) and Zambrano (2021b).

**Figure 2: Trend interruption estimates, food imports**

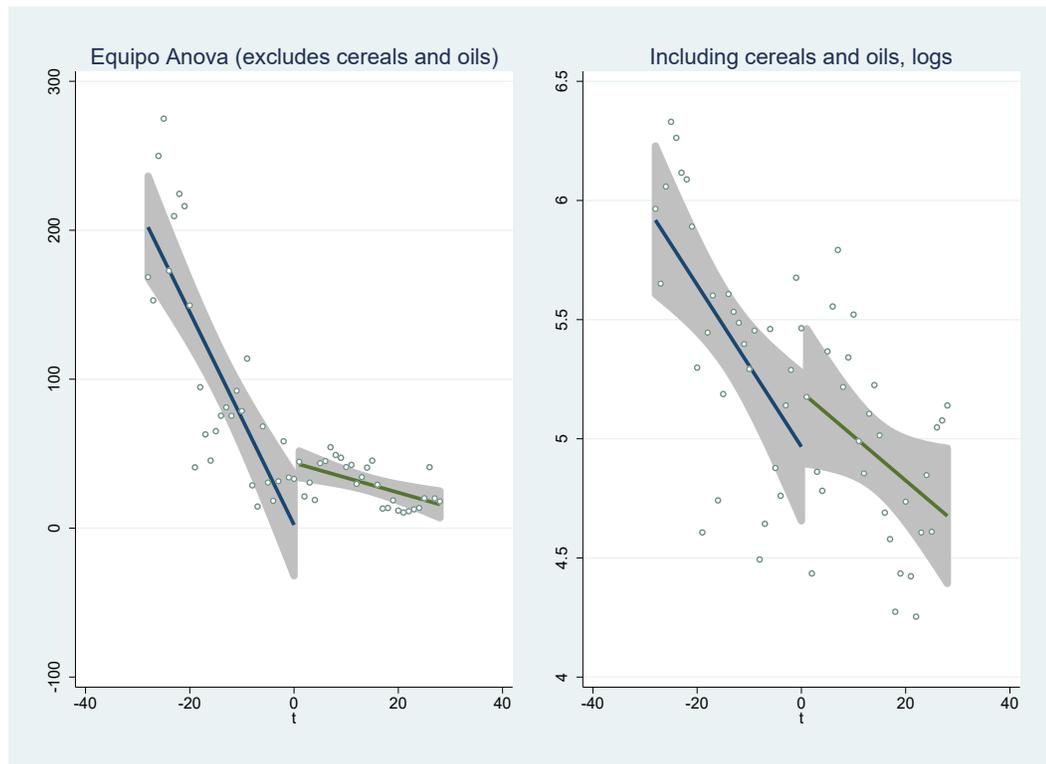

Sources: UN Comtrade, own calculations

The last column of Table 2 shows the results for medicines imports. In this case, the estimated increase in import levels at the time of sanctions also disappears once we adopt a logarithmic specification (panels 2 and 3, column 3). In fact, the point estimate on the levels coefficient turns negative (albeit insignificant). The trend coefficient, in contrast, remains strongly positive, indicating an increase in the rate of growth of food imports.

Figure 3 considers this result in greater detail. Note that medicines imports were declining at a rate of 9 log points per month, or the equivalent of a 66% decline per year, before the 2017 sanctions. Since this decline was strongly statistically significant (t=12.6), just a stabilization of imports at very low levels would be enough to result in a statistically significant break in trend. That is what the data shows happened. In the post-sanctions sample, the trend coefficient is not significantly different from zero (t=1.3). In other words, the change in trend result essentially captures the fact that medicines imports stabilized at very low levels. The positive – though insignificant - post-sanctions slope coefficient appears to be mostly driven by observations from the last months of 2019, when the economy had begun to experience some recovery in oil output, rather than anything that happened in the vicinity of the imposition of the 2017 financial sanctions.[13]

**Figure 3: Trend interruption estimates, medicine imports (logs)**

---

[13] Venezuela's oil output hit a low of 644 tbd in September of 2019 and rose by 14% in the last three months of 2019, which also correspond to the last three months of the Equipo Anova sample and the highest post-sanctions levels of medicines imports.

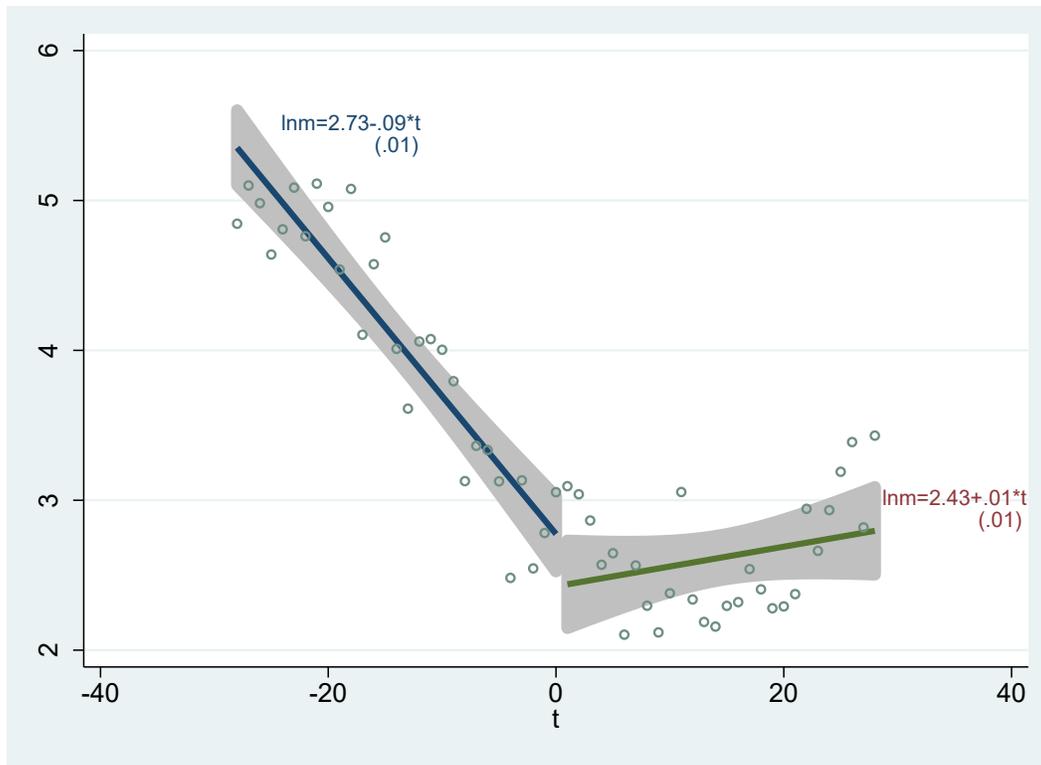

Sources: UN Comtrade, own calculations

Here it makes sense to think again in terms of implicit counterfactuals. Had medicines imports continued to decline at the pre-intervention rate of 9 log points per month after August of 2017, they would have fallen from $21 million per month at the time of sanctions (and $123 million per month two years before sanctions) to $1.5 million per month at the end of the sample in December 2019. While the logarithmic specification ensures that the projection of medicines imports at the end of the sample is greater than zero and thus mathematically feasible, that does not mean that it captures a realistic counterfactual of what would have happened to medicines imports in the absence of sanctions. Put differently, what the Equipo Anova estimate for medicines tells us is simply that medicines imports did not continue shrinking at a rate of two-thirds per year. It in no way establishes that it is reasonable to think of a sustained rate of decline to near-zero import levels as a reasonable counterfactual scenario in the absence of sanctions.

An extensive empirical literature shows that medicines demand is income inelastic, implying that the share of income devoted to medicines purchases increases as incomes drop (Danzon et al., 2015, Ringel et al., 2005; Farag et al., 2012).[14] Therefore, a country that suffers a sustained decline in incomes like that of Venezuela should see a relative stabilization of medicines demand with the decline in imports, even in the absence of changes in public policies. In other words, even the logarithmic specification may be insufficient to capture nonlinearities in the counterfactual expected path of the economy in the presence of income-inelastic demand for medicines.[15] Furthermore, the sole existence of economic and political elites with access to sources of foreign exchange suggests that they will always find a way to ensure their access to basic medicines, implying that we should expect medicines import demand to be bounded above zero.

---

[14] Generally, most values in the literature tend to be between 0.6 and 0.9. See Fan and Savedoff, 2012; Gerdtham and Jönsson 2000; Baltagi and Moscone 2010; Xu et al. 2011; Lépine, 2014; Tangtipongkul, 2016; Jeetoo and Jaunky, 2022; Dubey, 2020; Magsi et al, 2021.

To a certain extent, it is precisely the concern with the sensitivity to functional form of trend break specifications that has led to the development of an extensive literature in RDD approaches. While reiterating our prior caveats on the application of RDD designs to low-frequency data and in the absence of information on covariates to assess validity, it is instructive to consider the results of applying a more conventional RDD framework instead of the linear trend break specification to this data. We show these results in Table 4, where we test for the existence of breaks in trend with a more appropriate RDD design that uses local polynomial estimators with robust bias-corrected intervals, a triangular kernel and a mean-squared error optimal bandwidth selector as suggested by Calonico et al. (2017, 2020). In order to appropriately select the bandwidths, we use data from January 2012 to December 2020. We note that the optimal bandwidths selected by this method range from 8.5 to 12.9 months, well below the 28 months selected by Equipo Anova (2021).

**Table 4: Regression Discontinuity Design Estimates**

| | | ANOVA food imports (excluding cereals and oils) | Food imports (including cereals and oils) | Medicine imports |
|---|---|---|---|---|
| Regression Discontinuity, Logarithms | Change in level | -0.14 (0.39) | -0.75 (0.58) | 0.68*** (0.18) |
| | Change in slope | -0.55* (0.32) | -0.82* (0.45) | -0.23* (0.16) |

We find no evidence of an improvement in food and medicine imports in the RDD estimates associated with the imposition of financial sanctions. Regarding food imports, both the level and the slope of the trend *decline* around the time of sanctions, with the change in slope being borderline significant regardless of whether we include cereals and oils. Thus, if anything the data points in the direction of suggesting that a *deterioration* in availability of food imports occurred at the time of sanctions. In the case of medicine imports, the level and slope coefficient estimates point in different directions, with the point estimates indicating a strongly significant increase in the level of imports at the time of sanctions and a borderline significant decline in the slope of the trend. This suggests that there may have been a short-lived rise in medicines imports at the time of sanctions (possibly driven by rising oil prices in the fourth quarter of 2017). The magnitude of this increase is around three times the change in slope, suggesting that any improvement in medicines import levels was offset in less than four months by the increased rate of decline over time.

It may seem hard to reconcile the medicines RDD results (which find an increase in levels yet no change in trends) with those from the trend break specification (which find a stabilization in trend but no significant change in levels). Yet a close look at the data in Figure 3 shows that they are in fact compatible. The RDD result weighs heavily observations at the time of the 2017 financial sanctions, and we can see that these did in fact increase for a few months after the sanctions, but then declined steeply, only to begin to rise again at the end of 2019. In contrast, the trend break estimator in equation (1) finds no statistically significant trend in the post-sanctions data precisely because it is influenced both by observations near the cutpoint (which indicate a decline) as well as those far from it (which show some recovery).

---

[15] The issue is not just one of functional form, but rather of the incompleteness of the univariate framework to adequately assess the effect of an intervention such as sanctions when many other relevant covariates are changing strongly. A more adequate framework for assessing this question could be to ask the extent to which medicines purchases deviated from those that would have been expected given the country's economic contraction and existing income elasticity estimates. Exploring this and other alternative frameworks is a possible direction for future research.

### 4. Comments on replication

As noted previously, Equipo Anova has not made replication data or files available for their study, nor did they reply to our e-mail requests for these. While their data is constructed from a public United Nations Trade Statistics Comtrade site, the data in that site changes over time as participant countries update their statistics. This makes it impossible to reproduce the exact data set used by the Equipo Anova team.

We approach this problem by attempting to replicate as closely as possible the Comtrade data available at time of writing of the Equipo Anova study to reconstruct their series. Although Comtrade does not maintain a version of their data set at different moments in the past, they do publish the date and time of submission of each country-level observation and its updates. Using this information, we constructed a proxy of the data available at the time of writing of the Equipo Anova report by deleting observations for which no data had been submitted by the partner country at that time. Note that in doing so we kept observations for which the trading partner had submitted some data at the time, even if that data was subsequently updated (we can only observe the updated data). The logic of this is that we consider the updated observation (which is available to us) a better proxy than zero for the observation available at time of their writing (which is not available to us.) Zero is, of course, the value that would be implicitly attributed to the observation were we to omit it from our aggregate estimate. Since we do not know with precision the time at which Equipo Anova downloaded the data used in their report (published in January of 2021), we experimented with several dates in 4Q20 and chose October 1, 2020 as the estimated download time given that it made our estimated series closest to the one shown in their paper. This calculation was done with the Comtrade data available as of August 2022.

Ideally, we would want to compare our reconstructed series with the Equipo Anova data to see how closely we have replicated their findings. While we do not have the Equipo Anova series, we do have the scatter plots of the variables against time and the regression coefficient estimates published in their paper. We thus used Plot Digitizer, a data extraction tool that converts visual graphs to numerical series, to infer their data from their published figures. The extracted data replicates the coefficient estimates reported by Equipo Anova in their tables to a precision of less than one decimal point. The reconstructed and the extracted series have correlation coefficients of .9998 for food and .9993 for medicines, while the point estimates, standard errors and pre and post-sanctions average levels are nearly identical (Table 5). We reach the conclusion that Equipo Anova excluded commodity categories 10-19 from their food imports series based on observing that only the reconstructed series excluding these categories matches both their published results and the data extracted from their images.[16]

**Table 5: Comparison of results from Equipo Anova scatter plots and Comtrade replication.**

| | | Data extracted from Equipo Anova scatter plots | Data replicated with Comtrade data |
|---|---|---|---|
| Equipo Anova food imports (excluding cereals and oils) | Change in level | 44.26*** (13.68) | 44.86*** (13.82) |
| | Change in slope | 6.48*** (1.09) | 6.46*** (1.09) |
| | Average level | 66.29 | 66.37 |
| | Pre-sanctions | 104.94 | 104.60 |
| | Post-sanctions | 28.98 | 29.47 |
| | Correlation | 0.9998 | |
| Medicine imports | Change in level | 21.47*** (5.76) | 19.6*** (5.70) |
| | Change in slope | 5.79*** (0.52) | 5.47*** (0.49) |
| | Average level | 44.96 | 44.98 |
| | Pre-sanctions | 80.84 | 80.43 |
| | Post-sanctions | 10.32 | 10.75 |
| | Correlation | 0.9992 | |

## 5. Concluding remarks

In contrast to the claims of Equipo Anova (2021), we find no evidence of the improvement in imports of food and medicines at the time at which the U.S. imposed financial sanctions on Venezuela in August 2017. Instead, we find their results to be a consequence of data coding errors and questionable methodological choices. Their research design uses an unreasonable functional form that implies a counterfactual of negative imports in the absence of sanctions and omits data accounting for four-fifths of the country's food imports at the time of sanctions. Once these errors are corrected, any evidence of an improvement in the level or rate of change in food imports disappears. While we do find that medicines imports stabilized at very low levels after sanctions, the most reasonable explanation for this result appears to be that it is a consequence of the low income elasticity of medicines demand. Correct application of the regression discontinuity design methods which Equipo Anova (2021) claim incorrectly to have used delivers no evidence of a consistent improvement in imports of essentials, with most of the point estimates being of the opposite sign to that claimed by Equipo Anova. Neither close inspection of the corrected data nor a battery of statistical tests shows evidence of any sustained significant improvement in food or medicines imports following the 2017 financial sanctions.

---

[16] As shown in Table 5, the food series extracted from the figures published by Equipo Anova (2021) has a .9998 correlation and an average value that is nearly identical to that of the reconstructed series generated excluding categories 10-19. In contrast, it has a .8841 correlation and an average value that is one-third of the average level of the reconstructed series generated including categories 10-19.

## 6. References

Afesorgbor, S. K., and Mahadevan, R. (2016). The Impact of Economic Sanctions on Income Inequality of Target States. *World Development*, 83, 1–11.

Allen, S. and Lektzian, D. (2012). Economic sanctions: A blunt instrument? *Journal of Peace Research*, 50(1).

Amaya, V. (2021). Maduro, las sanciones y la mutación en la economía venezolana. Tal Cual. February, 10. https://talcualdigital.com/maduro-las-sanciones-y-la-mutacion-en-la-economia-venezolana/

Armando Armas, Twitter post, July 2022, 9:25 am, https://twitter.com/ArmandoArmas/status/1546485743274565633?s=20andt=SBSqM1e6QgqEzG4DuipPMA

Bahar, D., Bustos, S., Morales, J. R., and Santos, M. A. (2019) Impact of the 2017 sanctions on Venezuela: Revisiting the evidence. Brookings Institution. https://www.brookings.edu/wp-content/uploads/2019/05/impact-of-the-2017-sanctions-on-venezuela_final.pdf

Baltagi, B. and Moscone, F. (2010). Health care expenditure and income in the OECD reconsidered: Evidence from panel data. *Economic Modelling*, 27(4): 804-811.

Batmanghelidj, E. (2022). The Inflation Weapon: How American Sanctions Harm Iranian Households. Fourth Freedom Forum. Sanctions and Security Research Project. https://sanctionsandsecurity.org/publications/the-inflation-weapon-how-american-sanctions-harm-iranian-households/

Calonico, S., Cattaneo, M., Farrell, H., and Titiunik, R. (2017). rdrobust: Software for regression discontinuity designs. *Stata Journal* 17, 372–404

Calonico, S., Cattaneo, M., and Farrell, H. (2020). Optimal bandwidth choice for robust bias-corrected inference in regression discontinuity designs. *Econometrics Journal*, 23: 192–210.

Carreras, M., Visconti, G. and Acácio I. (2021). THE TRUMP ELECTION AND ATTITUDES TOWARD THE UNITED STATES IN LATIN AMERICA. *Public Opinion Quarterly*, 00(0): 1 – 11.

Carter Center (2022). Syria: From Punitive Sanctions to an Incentive-Based Approach. Fourth Freedom Forum. Sanctions and Security Research Project. https://sanctionsandsecurity.org/publications/syria-from-punitive-sanctions-to-an-incentive-based-approach/

Choi, S. W., and Luo, S. (2013). Economic Sanctions, Poverty, and International Terrorism: An Empirical Analysis. *International Interactions*. 39(2): 17–245.

Cunningham, S. (2021). *Causal Inference: The Mixtape*, New Haven: Yale University Press, pp. 252-282.

Dai, M., Felbermayr, G., Kirilakha, A., Syropoulos, C., Yalcin, E., and Yotov, Y. (2021) Timing the Impact of Sanctions on Trade. School of Economics Working Paper Series 2021-7, LeBow College of Business, Drexel University.

Dubey, J. (2020). Income elasticity of demand for health care and its change over time: Across the income groups and levels of health expenditure in India. Working Papers 20/324, National Institute of Public Finance and Policy.

Equipo Anova (2021). Impacto de las Sanciones Financieras Internacionales contra Venezuela: Nueva Evidencia, 3(1): 1 – 3.

Fan, V. and Savedoff, W. (2012). The Health Financing Transition: A Conceptual Framework and Empirical Evidence. Results for Development Institute, Working Paper.

Farzanegan, M. R., Khabbazan, M. M., and Sadeghi, H. (2016). Effects of Oil Sanctions on Iran's Economy and Household Welfare: New Evidence from A CGE Model. Economic Welfare and Inequality in Iran.

Felbermayr, G.; Syropoulos, S.; Yalcin, E. and Yotov, Y. (2020). On the Heterogeneous Effects of Sanctions on Trade and Welfare: Evidence from the Sanctions on Iran and a New Database, School of Economics Working Paper Series 2020-4, LeBow College of Business, Drexel University.

Filmer, D. and Prtichett, L. (1999). Child Mortality and Public Spending on Health: How Much Does Money Matter? Policy Research Working Paper Series 1864, World Bank.

Gharehgozlia, O. (2017). An Estimation of the Economic Cost of Recent Sanctions on Iran Using the Synthetic Control Method. *Economic Letters*, 157.

Gerdtham, U. and Jönsson, B. (2000) International Comparisons of Health Expenditure: Theory, Data and Econometric Analysis. In: Culyer, A.J. and Newhouse, J.P., Eds., *Handbook of Health Economics*, 1A, 11-53.

Godard, M. Koning, P. and Lindeboom, M. (2019). Targeting Disability Insurance Applications with Screening. IZA Institute of Labor Economics, Discussion Paper Series, IZA DP No. 12343.

Guaidó, J. (2022a). Juán Guaidó presenta informe de gestión. Press release. VPI Tv. https://www.youtube.com/watch?v=ubh1zyTbXTA

Guaidó, J. (2022b). Juan Guaidó: "Nicolás Maduro ha creado una economía clandestina y mafias en Venezuela". El Clarin. June, 11. https://www.clarin.com/mundo/juan-guaido-nicolas-maduro-creado-economia-clandestina-mafias-venezuela-_0_KjKhcccJrL.html

Gutmann, J., Neuenkirch, M. and Neumeier, F. (2017) Sanctioned to Death? The Impact of Economic Sanctions on Life Expectancy and its Gender Gap. ILE Working Paper Series, 11. 2 -36.

Gutmann, J., Neuenkirch, M., Neumeier, F. and Steinbach, A. (2018). Economic Sanctions and Human Rights: Quantifying the Legal Proportionality Principle. Universitat Trier Research Papers in Economics No. 2/18.

Gutmann, J., Neuenkirch, M. & Neumeier, F. (2021). The Economic Effects of International Sanctions: An Event Study. CESifo Working Paper No. 9007.

Ha, L. and Nam, P. (2022). An investigation of relationship between global economic sanction and life expectancy: do financial and institutional system matter? *Development Studies Research*, 9(1):48 – 66.

Hahn, J., Todd, P. and Van der Klaauw, W. (2001), Identification and Estimation of Treatment Effects with a Regression-Discontinuity Design. Econometrica, 69: 201-209.

Hausman, C. and Rapson, D. (2018). Regression Discontinuity in Time: Considerations for Empirical Applications. *Annual Review of Resource Economics*, 10: 533 – 552.

Hausmann, R. and Rigobón, R. (2002). An Alternative Interpretation of the 'Resource Curse': Theory and Policy Implications. National Bureau of Economic Research Working Papers 9424.

Hausmann, R., and Muci, F. (2019). Don't Blame Washington for Venezuela's Oil Woes: A Rebuttal. Americas Quarterly. May 1. https://www.americasquarterly.org/article/dont-blame-washington-for-venezuelas-oil-woes-a-rebuttal/

Hejazi, J. and Emamgholipour S. (2022). The Effects of the Re-imposition of US Sanctions on Food Security in Iran. *International Journal of Health Policy Management*, 11(5): 651 - 657.

Hufbauer, G., Schott, J., Elliott, K. and Oegg, B. (2007). *Economic Sanctions RECONSIDERED*, 3rd edition, Peterson Institute for International Economics, Washington D. C., United States.

Imbens, G. and Lemieux, T. (2007). Regression Discontinuity Designs: A Guide to Practice. National Bureau of Economic Research, Working Papers 13039.

Iturbe, R. (2021). Estados Unidos quiere negociar con Maduro las sanciones a cambio de elecciones libres. ALNavío. February 14. https://alnavio.es/estados-unidos-quiere-negociar-con-maduro-las-sanciones-a-cambio-de-elecciones-libres/

Jeetoo, J. and Jaunky, V. (2022). An Empirical Analysis of Income Elasticity of Out-of-Pocket Healthcare Expenditure in Mauritius. *Healthcare*, 10(1): 1 – 21.

Kholodilin, K. and Netšunajev, A. (2018). Crimea and punishment: the impact of sanctions on Russian economy and economies of the euro area. *Baltic Journal of Economics*, 19(1): 39-51

Kim, Y. (2019). Economic sanctions and child HIV. *The International Journal of Health Planning and Management*, 34(2): 693 – 700.

Kim, M. (2019). Unintended Consequences of Health Care Reform in South Korea: Evidence from a Regression Discontinuity in Time Design. *SSRN Electronic Journal*.

Lee, D. and Lemieux, T. (2010). Regression Discontinuity Designs in Economics. *Journal of Economic Literature,* 48(2): 281 – 355.

Lee, M. (2016). *Matching, Regression Discontinuity, Difference in Differences, and Beyond.* Oxford University Press, 1st ed.

Lépine, A. (2014). Is Health a Necessity in Sub-Saharan Africa? An Investigation of Income-Elasticity of Health Expenditures in Rural Senegal. *Journal of International Development*, 27(7): 1153-1177.

Lovett, N. and Xue, Y. (2022). Rare homicides, criminal behavior, and the returns to police labor. *Journal of Economic Behavior & Organization*. 194, 172 – 195.

Lucena, C. and Apolinário, L. (2016) Targeted Versus Conventional Economic Sanctions: What Is at Stake for Human Rights? *International Interactions*. 42(4): 565–589.

Magsi, H., Memon, M., Sabir, M., Magsi, I. and Anwar, N. (2021). Income Elasticity of Household's health and Wellness in Rural Pakistan. *Journal of Economics and Management Sciences*, 2(1): 67-78.

Mendez, B. (2021). ANOVA: Sanciones sí tuvieron impacto en la producción de crudo, pero fueron positivas para las importaciones de alimentos y medicinas. Descifrado. January 22. https://www.descifrado.com/2021/01/22/anova-sanciones-si-tuvieron-impacto-en-la-produccion-de-crudo-pero-fueron-positivas-para-las-importaciones-de-alimentos-y-medicinas/

Morteza, G. (2021). Who is afraid of sanctions? Themacroeconomic and distributional effectsof the sanctions against Iran. *Wiley Economics and Politics*, 34(3): 395 – 428.

Neuenkirch, M. and Neumeier, F. (2015). The impact of UN and US economic sanctions on GDP growth. *European Journal of Political Economy,* 40A, 110–125.

Neuenkirch, M., and Neumeier, F. (2016). The impact of US sanctions on poverty. *Journal of Development Economics,* 121: 110–119.

Oliveros, L. (2020). Impacto de las sanciones financieras y petroleras en la economía venezolana. WOLA, Washington, DC. https://www.wola.org/wp-content/uploads/2020/10/Oliveros-Resumen-FINAL.pdf

Omar Zambrano, Twitter Post, January 2021a, 8:23 pm. https://twitter.com/Econ_Vzla/status/1352774002436165632?s=20andt=rPvPqzup2T-YV9kPqFKtSw
Omar Zambrano, Twitter Post, January 2021b, 8:23 pm. https://twitter.com/Econ_Vzla/status/1352774002436165632?s=20andt=rPvPqzup2T-YV9kPqFKtSw

Parker, D., Foltz, J., and Elsea, D. (2017). Unintended Consequences of Sanctions for Human Rights: Conflict Minerals and Infant Mortality. *Journal of Law & Economics*, 59(4): 731 – 774.

Peksen, D. (2011). Economic Sanctions and Human Security: The Public Health Effect of Economic Sanctions. *Foreign Policy Analysis,* 7(3): 237–251.

Peksen, D., and Drury, A. C. (2010). Coercive or Corrosive: The Negative Impact of Economic Sanctions on Democracy. *International Interactions,* 36(3): 240–264.

Peksen, D. (2009). Better or Worse? The Effect of Economic Sanctions on Human Rights. Journal of Peace Research, 46(1), 2009, pp. 59–77.

Petrescu, I (2016). The Humanitarian Impact Of Economic Sanctions. *Europolity* – Continuity and Change in European Governance - New Series, Department of International Relations and European Integration, National University of Political Studies and Public Administration, 10(2): 1-41.

Política UCAB (2021). ANOVA presenta estudio sobre nueva evidencia del impacto de las sanciones financieras internacionales contra Venezuela. Política UCAB. January, 21. https://politikaucab.net/2021/01/21/anova-presenta-estudio-sobre-nueva-evidencia-del-impacto-de-las-sanciones-financieras-internacionales-contra-venezuela/

Pritchett, L. and Summers, L. (1996). Wealthier is Healthier. *Journal of Human Resources*, 31(4): 841 – 868.

Ringel, J., Hosek, S. Vollaard, B. and Mahnovski, S. (2005). *The Elasticity of Demand for Health Care A Review of the Literature and Its Application to the Military Health System.* California, CA: RAND Corporation.

Rodríguez, F. and Hausmann, R. (2012). *Venezuela Before Chávez: Anatomy of an Economic Collapse.* Pennsylvania, PA: The Pennsylvania State University Press.

Rodríguez, F. and Sachs, J. (1999). Why Do Resource-Abundant Economies Grow More Slowly? *Journal of Economic Growth*, 4(3): 277–303.

Rodríguez, F. (2018). Crude Realities: Understanding Venezuela's Economic Collapse. WOLA. https://www.venezuelablog.org/crude-realities-understanding-venezuelas-economic-collapse/

Rodríguez, F. (2019). Sanctions and the Venezuelan economy: What the data say. Torino Economics.

Rodríguez, F. (2021). Scorched Earth: The Political Economy of Venezuela's Collapse, 2013-2020. Unpublished book manuscript.

Rodríguez, F. (2022a). Sanctions, Economic Statecraft, and Venezuela's Crisis Case Study. Fourth Freedom Forum, Sanctions and Security Research Project. https://sanctionsandsecurity.org/wp-content/uploads/2022/01/January-2022-Venezuela-Case_Rodriguez.pdf

Rodríguez, F. (2022b). The Human Consequences of Economic Sanctions. Working Paper.

Rodríguez, F. (2022c). The economic determinants of Venezuela's hunger crisis. MPRA Paper No. 113669.

Rodríguez, F. (2022d). Sanctions and oil production: Evidence from Venezuela's Orinoco Basin. *Latin American Economic Review* 31(6): 1-31.

Splinter, M. and Klomp, J. (2021). Do Sanctions Cause Economic Growth Collapses? *Netherlands Annual Review of Military Studies,* 115–132.

Tangtipongkul, K. (2016). Income Elasticity for Medical Care Services: An Empirical Study in Thailand. Thammasat. *Review of Economic and Social Policy*, 2(1): 76–123.

UNDP (United Nations Development Programme), (2010). *Human Development Report 2010: The Real Wealth of Nations: Pathways to Human Development.* New York, NY.

Van Bergeijk, P. (2015). Sanctions against Iran - A preliminary economic assessment. European Union Institute for Security Studies.

Warburton, C. (2016). The International Law and Economics of Coercive Diplomacy: Macroeconomic effects and empirical findings. *Applied Econometrics and International Development,* 16(1).

Wen, J., Zhao, X., Wang, Q. and Chang, C. (2020). The impact of international sanctions on energy security. Energy & Environment, 0(0): 1 – 23.

Wood, R. (2008). "A Hand upon the Throat of the Nation": Economic Sanctions and State Repression, 1976–2001. *International Studies Quarterly*, 52(3): 489–513.

Xu, K., Saksena, P., and Holly, A. (2011). The Determinants of Health Expenditure: A Country-Level Panel Data Analytis. Working Paper, Washington DC.